\documentstyle[11pt,epsf]{article}
\input epsf.tex
 1
 1
 1

\def\lsim{\mathrel{\raise.3ex\hbox{$<$\kern-.75em\lower1ex\hbox{$\sim$}}}}
\def\gsim{\mathrel{\raise.3ex\hbox{$>$\kern-.75em\lower1ex\hbox{$\sim$}}}}
\catcode`\@=11
\def\slash{\mathpalette\make@slash}
\def\make@slash#1#2{\setbox\z@\hbox{$#1#2$}%
  \hbox to 0pt{\hss$#1/$\hss\kern-\wd0}\box0}
\catcode`\@=12 

\begin{document}
\noindent
\thispagestyle{empty}
\renewcommand{\thefootnote}{\fnsymbol{footnote}}
\begin{flushright}
{\bf UCSD/PTH 98-01}\\
{\bf DESY 98-008}\\
{\bf hep-ph/9801397}\\
{\bf January 1998}\\
\end{flushright}
\vspace{.5cm}
\begin{center}
  \begin{Large}\bf
Top Quark Pair Production at Threshold: \\[2mm]
Complete Next-to-next-to-leading Order Relativistic Corrections
  \end{Large}
  \vspace{1.5cm}

\begin{large}
 A.~H.~Hoang$^{a}$ and
 T. Teubner$^{b}$ 
\end{large}

\vspace{1.5cm}
\begin{it}
${}^a$ Department of Physics, University of California, San Diego,\\
   La Jolla, CA 92093-0319, USA\\[.5cm]
${}^b$  Deutsches Elektronen-Synchrotron DESY,\\
   D-22603 Hamburg, Germany
\end{it}

  \vspace{2.5cm}
  {\bf Abstract}\\
\vspace{0.3cm}

\noindent
\begin{minipage}{14.0cm}
\begin{small}
The complete next-to-next-to-leading order (i.e.\ ${\cal{O}}(v^2)$,
${\cal{O}}(v \alpha_s)$ and 
${\cal{O}}(\alpha_s^2)$) relativistic corrections to the total photon
mediated $t\bar t$ production cross section at threshold are presented
in the framework of nonrelativistic quantum chromodynamics.
The results are obtained using semi-analytic methods and
``direct matching''. The size of the next-to-next-to-leading order
relativistic corrections is found to be comparable to the size of the
next-to-leading order ones. 
\\[3mm]
\end{small}
\end{minipage}
\end{center}
\setcounter{footnote}{0}
\renewcommand{\thefootnote}{\arabic{footnote}}
\vspace{1.2cm}
%
%
%
\newpage
\noindent
%
Due to the large top quark mass, which allows for the decay channel
$t\to b W$, $t\bar t$ production at threshold in lepton pair
collisions offers the unique opportunity to study heavy
quark-antiquark bound state and near threshold dynamics using
perturbative QCD~\cite{Fadin1}. With this motivation in mind a
considerable 
number of studies has been carried out in the past in order to
calculate $t\bar t$ production
observables~\cite{Strassler1,Kwong1,Jezabek1,Sumino1,japaner} and 
explore their potential for measurements of the top quark mass\footnote{
Throughout this paper $M_t$ is understood as the top quark pole mass.
} 
$M_t$ and the strong coupling $\alpha_s$ at future experiments like
the LC (Linear Collider)~\cite{NLC} or the FMC (First Muon
Collider)~\cite{Berger1}. In
view of the high precision which might be achieved for the QCD
calculations as well as for $t\bar t$ production measurements even
relatively small effects coming from a light Higgs
boson~\cite{Strassler1,higgseffects} have been investigated. However, the
present day analyses only include QCD effects up to next-to-leading
order (NLO) in form of the one-loop corrections to the QCD
potential~\cite{Fischler1,Billoire1} and various
${\cal{O}}(\alpha_s)$ short-distance 
corrections. A complete next-to-next-to-leading order (NNLO)
calculation, which would be necessary to study the reliability of the
present day analyses and to make the consideration of small effects
from beyond QCD at all feasible, has been missing so far.

In this letter we present the complete NNLO relativistic corrections
to the total photon mediated $t\bar t$ production cross section. 
As NNLO we count all corrections of order $v^2$, $v\,\alpha_s$ and
$\alpha_s^2$ relative to the cross section in the nonrelativistic
limit, $v$ being the c.m.\ velocity of the top quarks.
As relativistic, on the other hand, we count the corrections coming
from the top quark kinetic energy, the $t\bar t$ production and
annihilation process including the short-distance corrections, and the
$t\bar t$ interaction potentials.
We use NRQCD~\cite{Caswell1} to conveniently parameterize
calculations 
and results in a systematic manner following the approach proposed
in~\cite{Hoang1}. The calculations are carried out using semi-analytic
methods and the ``direct matching'' procedure introduced
in~\cite{Hoang2}. We would like to point out that NNLO corrections
involving the top quark decay are not determined here. The latter
effects would include ${\cal{O}}(\alpha_s^2)$ two loop corrections to
the free top quark width and a consistent treatment of the effects
from the off-shellness of the decaying top quarks, the time dilatation
and the interactions among the decay products and the other top quark
(if it is not decayed yet). Although the size and the interplay of all
these effects have been studied at various places in the literature
(see e.g.\ \cite{Sumino1,Teubner1,Moedritsch1}), their consistent
treatment at NNLO still remains an open problem. As
far as the NNLO relativistic corrections discussed in this letter are
concerned we will use the naive replacement 
\begin{equation}
E\equiv\sqrt{s} - 2 M_t \longrightarrow \tilde E = E + i \, \Gamma_t
\label{replacement}
\end{equation}
in the spirit of~\cite{Fadin1} in order to examine their size and
properties, where $\Gamma_t$ represents a constant which
is not necessarily the decay width of a free top quark.
We also would like to emphasize that we treat {\it all interactions}
purely perturbatively and that nowhere in this work the confining
long-range contributions to the QCD potential or other nonperturbative
effects are taken into account. This is somewhat contrary to the
standard present day approach used to describe $t\bar t$ production at 
threshold (see
\cite{Strassler1,Kwong1,Jezabek1,Sumino1}), but we take
the position that nonperturbative effects might be added later as a
correction.
%
%

We start from the NRQCD Lagrangian
\begin{eqnarray}
\lefteqn{
{\cal{L}}_{\mbox{\tiny NRQCD}} \, = \, 
- \frac{1}{2} \,\mbox{Tr} \, G^{\mu\nu} G_{\mu\nu} 
+ \sum_{q=u,d,s,c,b} \bar q \, i \slash{D} \, q
+\, \psi^\dagger\,\bigg[\,
i D_t 
+ a_1\,\frac{{\mbox{\boldmath $D$}}^2}{2\,M_t} 
+ a_2\,\frac{{\mbox{\boldmath $D$}}^4}{8\,M_t^3}
\,\bigg]\,\psi + \ldots 
}\nonumber\\[2mm] & &
+ \,\psi^\dagger\,\bigg[\, 
\frac{a_3\,g_s}{2\,M_t}\,{\mbox{\boldmath $\sigma$}}\cdot
    {\mbox{\boldmath $B$}}
+ \, \frac{a_4\,g_s}{8\,M_t^2}\,(\,{\mbox{\boldmath $D$}}\cdot 
  {\mbox{\boldmath $E$}}-{\mbox{\boldmath $E$}}\cdot 
  {\mbox{\boldmath $D$}}\,)
+ \frac{a_5\,g_s}{8\,M_t^2}\,i\,{\mbox{\boldmath $\sigma$}}\,
  (\,{\mbox{\boldmath $D$}}\times 
  {\mbox{\boldmath $E$}}-{\mbox{\boldmath $E$}}\times 
  {\mbox{\boldmath $D$}}\,)
 \,\bigg]\,\psi 
+\ldots
\,.
\label{NRQCDLagrangian}
\end{eqnarray}
The gluonic and light quark degrees of freedom are described by the
conventional relativistic Lagrangian, whereas the top and antitop
quark are described by the Pauli spinors $\psi$ and $\chi$,
respectively. For convenience all color indices are suppressed. 
The straightforward antitop bilinears are omitted and
only those terms relevant for the NNLO cross section are displayed.
$D_t$ and {\boldmath $D$} are the time and space components of the
gauge covariant derivative $D_\mu$, and $E^i = G^{0 i}$ and $B^i =
\frac{1}{2}\epsilon^{i j k} G^{j k}$ the electric and magnetic
components of the gluon field strength tensor (in Coulomb gauge).
The short-distance coefficients $a_1,\ldots,a_5$ are normalized to one
at the Born level. Because we use ``direct matching''~\cite{Hoang2} 
their higher order contributions are irrelevant for this work. 

To formulate the normalized total $t\bar t$ production cross
section (via a virtual photon) 
$R = \sigma({e^+e^- \atop \mu^+\mu^-}\to\gamma^*\to t\bar t)/\sigma_{pt} $
($\sigma_{pt} =4\pi\alpha^2/3s$) in the nonrelativistic region at NNLO
in NRQCD we start from the fully covariant expression for the cross
section
\begin{equation}
R(q^2) \, = \,
\frac{4\,\pi\,Q_t^2}{q^2}\,\mbox{Im}[\,
\langle\,0\,|\, T\, \tilde j_\mu(q) \,\tilde j^\mu(-q)\, |\,0\,\rangle
\,]
\,,
\label{crosssectioncovariant}
\end{equation}
where $Q_t=2/3$ is the top quark electric charge.
We then expand the electromagnetic current (in momentum space)
$\tilde j_\mu(\pm q) = (\tilde{\bar t}\gamma^\mu \tilde t)(\pm q)$
which produces/annihilates a $t\bar t$ pair with c.m.\ energy
$\sqrt{q^2}$ in terms of ${}^3\!S_1$
NRQCD currents up to dimension eight ($i=1,2,3$)
\begin{equation}
\tilde j_i(q) = b_1\,\Big({\tilde \psi}^\dagger \sigma_i 
\tilde \chi\Big)(q) -
\frac{b_2}{6 M_t^2}\,\Big({\tilde \psi}^\dagger \sigma_i
(\mbox{$-\frac{i}{2}$} 
\stackrel{\leftrightarrow}{\mbox{\boldmath $D$}})
 \tilde \chi\Big)(q) + \ldots
\,,
\label{currentexpansion}
\end{equation}
where the constants $b_1$ and $b_2$ are short-distance coefficients
normalized to one at the Born level. 
The expansion of $\tilde j_i(- q)$ is obtained from
Eq.~(\ref{currentexpansion}) via charge conjugation symmetry. 
It should be noted that only the spatial components of the currents
contribute. 
Inserting expansion~(\ref{currentexpansion}) back into
Eq.~(\ref{crosssectioncovariant}) leads to the NRQCD expression
of the nonrelativistic cross section at the NNLO level
\begin{eqnarray}
R_{\mbox{\tiny NNLO}}^{\mbox{\tiny thr}}(\tilde E) & = &
\frac{\pi\,Q_t^2}{M_t^2}\,C_1(\mu_{\rm hard},\mu_{\rm fac})\,
\mbox{Im}\Big[\,
{\cal{A}}_1(\tilde E,\mu_{\rm soft},\mu_{\rm fac})
\,\Big]
\nonumber\\[2mm] & & 
- \,\frac{4 \, \pi\,Q_t^2}{3 M_t^4}\,
C_2(\mu_{\rm hard},\mu_{\rm fac})\,
\mbox{Im}\Big[\,
{\cal{A}}_2(\tilde E,\mu_{\rm soft},\mu_{\rm fac})
\,\Big]
+ \ldots
\,,
\label{crosssectionexpansion}
\end{eqnarray}
where
\begin{eqnarray}
{\cal{A}}_1 & = & \langle \, 0 \, | 
\, ({\tilde\psi}^\dagger \vec\sigma \, \tilde \chi)\,
\, ({\tilde\chi}^\dagger \vec\sigma \, \tilde \psi)\,
| \, 0 \, \rangle
\,,
\label{A1def}
\\[2mm]
{\cal{A}}_2 & = & \mbox{$\frac{1}{2}$}\,\langle \, 0 \, | 
\, ({\tilde\psi}^\dagger \vec\sigma \, \tilde \chi)\,
\, ({\tilde\chi}^\dagger \vec\sigma \, (\mbox{$-\frac{i}{2}$} 
\stackrel{\leftrightarrow}{\mbox{\boldmath $D$}}) \tilde \psi)\,
+ \mbox{h.c.}\,
| \, 0 \, \rangle
\,.
\label{A2def}
\end{eqnarray}
To obtain the factor $4 \pi Q_t^2/3 M_t^4$ in the second line of
expression~(\ref{crosssectionexpansion}) we have already used
relation~(\ref{A2Greenfunctionrelation}).
The cross section is expanded in terms of a sum of absorptive parts of
nonrelativistic current correlators (containing long-distance
physics\footnote{
In the context of this paper ``long-distance'' is not equivalent to
``nonperturbative''.
}) 
multiplied by short-distance coefficients $C_i$
($i=1,2,\ldots$). In Eq.~(\ref{crosssectionexpansion})
we have also shown the dependences on the various renormalization
scales: the soft scale $\mu_{\rm soft}$ and the hard scale $\mu_{\rm hard}$ are
governing the perturbative expansions of the correlators and the
short-distance coefficients, respectively, and arise from the light
degrees of freedom in ${\cal{L}}_{\mbox{\tiny NRQCD}}$,\footnote{
Throughout this work we use the convention 
$\alpha_s = \alpha_s^{(n_l=5)}$
in the $\overline{\mbox{MS}}$ scheme.
}
whereas the factorization scale $\mu_{\rm fac}$ essentially
represents the boundary between hard (i.e.\ of order $M_t$) and soft
momenta. This boundary is not uniquely defined and therefore both,
the correlators and the short-distance coefficients, in general depend
on $\mu_{\rm fac}$ (leading to new anomalous dimensions). Because the
term in the second line in Eq.~(\ref{crosssectionexpansion}) is
already of NNLO 
(i.e.\ suppressed by $v^2$) we can set $C_2=1$ and ignore the
factorization scale dependence of the correlator ${\cal{A}}_2$.
The calculation of all terms in
expression~(\ref{crosssectionexpansion}) proceeds in two basic
steps.\\[2mm]
{\it Step 1: Calculation of the nonrelativistic correlators.} -- The
calculation of the correlators at NNLO is simplified enormously by the
fact that for the $t\bar t$ pair, which is produced and annihilated in a
color singlet S-wave state, all retardation effects can be
neglected. Technically this means that the transverse
gluon exchange, which would lead to temporally retarded interactions
(closely related to Lamb-shift type corrections known in QED), can be
treated as instantaneous, i.e.\ its energy dependence can be
ignored. This can be seen by using formal counting rules (see
e.g.\ \cite{Labelle1,Grinstein1}) or from physical considerations
because a real gluon (i.e.\ one that carries energy) which is radiated
after the $t\bar t$ pair was produced has to be absorbed before
the $t\bar t$ pair is annihilated. Phase space effects and the dipole
matrix element then lead to a suppression of this process $\propto
v^3$, i.e.\ to an effect beyond NNLO. In other words,
as far as the nonrelativistic correlators in
Eq.~(\ref{crosssectionexpansion}) are concerned, NRQCD reduces to a
two-body (top-antitop) Schr\"odinger theory. The potentials in the
resulting Schr\"odinger equation are determined by considering $t\bar
t\to t\bar t$ one gluon exchange t-channel scattering amplitudes in
NRQCD. To NNLO 
(i.e.\ including potentials suppressed by at most $\alpha_s^2$,
$\alpha_s/M_t$ or
$1/M_t^2$ relative to the Coulomb potential) the relevant potentials
read ($a_s\equiv\alpha_s(\mu_{\rm soft})$, $C_A=3$, $C_F=4/3$,
$T=1/2$, $\tilde\mu\equiv e^\gamma\,\mu_{\rm soft}$, $r\equiv | \vec r
|$) 
\begin{eqnarray}
V_c(\vec r) & = & -\,\frac{C_F\,a_s}{r}\,
\bigg\{\, 1 +
\Big(\frac{a_s}{4\,\pi}\Big)\,\Big[\,
2\,\beta_0\,\ln(\tilde\mu\,r) + a_1
\,\Big]
\nonumber\\[2mm] & & \quad
 + \Big(\frac{a_s}{4\,\pi}\Big)^2\,\Big[\,
\beta_0^2\,\Big(\,4\,\ln^2(\tilde\mu\,r) 
      + \frac{\pi^2}{3}\,\Big) 
+ 2\,\Big(2\,\beta_0\,a_1 + \beta_1\Big)\,\ln(\tilde\mu\,r) 
+ a_2
\,\Big]
\,\bigg\}
\,,
\\[2mm]
V_{\mbox{\tiny BF}}(\vec r) & = & 
\frac{C_F\,a_s\,\pi}{M_t^2}\,
\Big[\,
1 + \frac{8}{3}\,\vec S_t \,\vec S_{\bar t}
\,\Big]
\,\delta^{(3)}(\vec r)
+ \frac{C_F\,a_s}{2\,M_t^2 r}\,\Big[\,
\vec\nabla^2 + \frac{1}{r^2} \vec r\, (\vec r \, \vec\nabla) \vec\nabla
\,\Big]
\nonumber\\[2mm] & &
- \,\frac{3\,C_F\,a_s}{M_t^2\,r^3}\,
\Big[\,
\frac{1}{3}\,\vec S_t \,\vec S_{\bar t} - 
\frac{1}{r^2}\,\Big(\vec S_t\,\vec r\,\Big)
\,\Big(\vec S_{\bar t}\,\vec r\,\Big)
\,\Big]
+ \frac{3\,C_F\,a_s}{2\,M_t^2\,r^3}\,\vec L\,(\vec S_t+\vec S_{\bar t})
\label{BFpotential}
\,,\\[2mm]
V_{\mbox{\tiny NA}}(\vec r) & = &
-\,\frac{C_A\,C_F\,a_s^2}{2\,M_t\,r^2} 
\,,
\label{NApotential}
\end{eqnarray}
where $\vec S_t$ and $\vec S_{\bar t}$ are the top and antitop spin
operators and $\vec L$ is the angular momentum operator and ($n_l=5$)
\begin{eqnarray}
\beta_0 & = & \frac{11}{3}\,C_A - \frac{4}{3}\,T\,n_l
\,,
\nonumber\\[2mm]
\beta_1 & = & \frac{34}{3}\,C_A^2 
-\frac{20}{3}C_A\,T\,n_l
- 4\,C_F\,T\,n_l
\,,
\nonumber\\[2mm]
a_1 & = &  \frac{31}{9}\,C_A - \frac{20}{9}\,T\,n_l
\,,
\nonumber\\[2mm]
a_2 & = & 
\bigg(\,\frac{4343}{162}+6\,\pi^2-\frac{\pi^4}{4}
 +\frac{22}{3}\,\zeta_3\,\bigg)\,C_A^2 
-\bigg(\,\frac{1798}{81}+\frac{56}{3}\,\zeta_3\,\bigg)\,C_A\,T\,n_l
\nonumber\\[2mm] & &
-\bigg(\,\frac{55}{3}-16\,\zeta_3\,\bigg)\,C_F\,T\,n_l 
+\bigg(\,\frac{20}{9}\,T\,n_l\,\bigg)^2
\,.
\end{eqnarray}
The constants $\beta_0$ and $\beta_1$ are the one- and two-loop
coefficients of the QCD beta function and $\gamma=0.577216\ldots$ is the
Euler constant. $V_c$ is the Coulomb (static) potential. Its
${\cal{O}}(\alpha_s)$ and ${\cal{O}}(\alpha_s^2)$ corrections have
been determined in~\cite{Fischler1,Billoire1} and \cite{Peter1},
respectively. 
$V_{\mbox{\tiny BF}}$ is the Breit-Fermi potential known from
positronium and $V_{\mbox{\tiny NA}}$ a purely non-Abelian potential generated
through non-analytic terms in one-loop NRQCD (or QCD) diagrams
containing the triple gluon vertex (see
e.g.\ \cite{nonabelianpotential} for older 
references).
The nonrelativistic correlators are directly related to the Green
function of the Schr\"odinger equation
\begin{equation}
\bigg( -\frac{\vec\nabla^2}{M_t} - \frac{\vec\nabla^4}{4M_t^3} +
V_{c}(\vec r) + V_{\mbox{\tiny BF}}(\vec r) + V_{\mbox{\tiny
NA}}(\vec r)  
-\tilde E\,\bigg)\,G(\vec r,\vec r^\prime,\tilde E)
\, = \, 
\delta^{(3)}(\vec r-\vec r^\prime) 
\,,
\label{Schroedingerfull}
\end{equation}
where $V_{\mbox{\tiny BF}}$ is evaluated for the ${}^3\!S_1$ configuration
only. The correlators read
\begin{eqnarray}
{\cal{A}}_1 & = & 6\,N_c\,\Big[\,
\lim_{|\vec r|,|\vec r^\prime|\to 0}\,G(\vec r,\vec r^\prime,\tilde E)
\,\Big]
\,,
\label{A1Greenfunctionrelation}
\\
{\cal{A}}_2  
& = & 
M_t\,\tilde E\,{\cal{A}}_1
\,.
\label{A2Greenfunctionrelation}
\end{eqnarray}
Relation~(\ref{A1Greenfunctionrelation}) can be easily inferred by
taking into account that the Green function $G(\vec r,\vec
r^\prime,\tilde E)$ describes the propagation of a top-antitop pair
which is produced and annihilated at distances $|\vec r|$ and $|\vec
r^\prime|$, respectively. 
Because the exact solution of Eq.~(\ref{Schroedingerfull}) seems to be
an impossible task, we rely on a numerical solution of the equation
\begin{equation}
\bigg( -\frac{\vec\nabla^2}{M_t} + V_c(\vec r)
-\tilde E\,\bigg)\,G_c(\vec r,\vec r^\prime,\tilde E)
\, = \, \delta^{(3)}(\vec r-\vec r^\prime)
\label{SchroedingerCoulomb}
\end{equation}
using techniques developed in~\cite{Strassler1,Jezabek1}.
The result for $G_c(0,0,\tilde E)$ is then combined with the
corrections to the leading order (LO) Coulomb Green function
$G_c^{(0)}$~\cite{Wichmann1} (defined 
through Eq.~(\ref{SchroedingerCoulomb}) for $V_c(\vec r) = -C_F a_s/r$) 
coming from the kinetic energy correction
and the potentials $V_{\mbox{\tiny BF}}$ and $V_{\mbox{\tiny NA}}$.
These corrections are calculated analytically using
Rayleigh-Schr\"odinger time-independent perturbation theory,
\begin{eqnarray}
\delta G(\vec r,\vec r^\prime,\tilde E) & = &
\int d\vec x^3\, G_c^{(0)}(\vec r,\vec x,\tilde E)\,
\Big[\,\frac{\vec\nabla^4}{4 M_t^3} 
- V_{\mbox{\tiny BF}}(\vec x) - V_{\mbox{\tiny NA}}(\vec x)\,\Big]
\,G_c^{(0)}(\vec x,\vec r^\prime,\tilde E)
\,.
\label{GreenfunctionBF}
\end{eqnarray}
For the calculation of Eq.~(\ref{GreenfunctionBF}) we use techniques
employed in~\cite{Hoang3}, where the Abelian NNLO contributions have
already been determined.
The final result for ${\cal{A}}_1$ at NNLO reads 
\begin{eqnarray}
\lefteqn{
{\cal{A}}_1  \, = \,
6\,N_c\,\bigg[\,
G_c(0,0,\tilde E) - G_c^{(0)}(0,0,\tilde E)
\,\bigg]
}
\label{A1explicit}
\\[2mm] & & 
+ \frac{N_c\,C_F\,a_s\,M_t^2}{2 \pi}\,
\bigg(\, 1 + \frac{3}{2}\,\frac{C_A}{C_F}
\,\bigg)\,
\bigg\{\,
i\,\tilde v - C_F\,a_s\,\bigg[\,
\ln(-i \frac{M_t\,\tilde v}{\mu_{\rm fac}}) + \gamma 
  + \Psi\bigg( 1-i\,\frac{C_F\,a_s}{2\,\tilde v} \bigg)
\,\bigg]
\,\bigg\}^2
\nonumber\\[2mm] & &
+\frac{3\,N_c\,M_t^2}{2\,\pi}\,
\bigg\{
i\,\tilde v\,\Big(1+\frac{5}{8}\,\tilde v^2\Big)
 - C_F\,a_s\,\Big(1+2\,\tilde v^2\Big) \bigg[
\ln(-i \frac{M_t\,\tilde v}{\mu_{\rm fac}}) + \gamma 
   + \Psi\bigg( 1-i\,\frac{C_F\,a_s\,
    (1+\frac{11}{8}\,\tilde v^2)}{2\,\tilde v} \bigg)
\bigg]
\bigg\}
\,,
\nonumber
\end{eqnarray}
where
\begin{equation}
\tilde v \, \equiv \, \sqrt{\frac{\tilde E}{M_t}}
\,,
\end{equation}
and $\Psi$ is the digamma function,
$\Psi(z)\equiv\frac{d}{dz}\ln\Gamma(z)$.
In the first line of Eq.~(\ref{A1explicit}) the LO Green function has
been subtracted to avoid double counting of the LO contribution
contained in the third line. It should be noted that the limit $|\vec
r|,|\vec r^\prime|\to 0$ in expression~(\ref{GreenfunctionBF}) causes
UV~divergences 
which are regulated using the short-distance cutoff
$\mu_{\rm fac}$. Further, all power divergences
$\propto \mu_{\rm fac}/M_t$ are subtracted~\cite{Bodwin1} and
$\mu_{\rm fac}$ is defined in a way that 
expression~(\ref{A1explicit}) takes the simple form shown above. The
corresponding result in any other regularization scheme could be
obtained from the one presented here through a redefinition of the
factorization scale.    
For ${\cal{A}}_2$ only the
LO contribution in Eq.~(\ref{A1explicit}) is relevant and we arrive at
\begin{eqnarray}
{\cal{A}}_2 & = & 
\tilde v^2\,\frac{3\,N_c\,M_t^4}{2\,\pi}\,
\bigg\{\,
i\,\tilde v - C_F\,a_s\,\bigg[\,
\ln(-i \frac{M_t\,\tilde v}{\mu_{\rm fac}}) + \gamma 
  + \Psi\bigg( 1-i\,\frac{C_F\,a_s}{2\,\tilde v} \bigg)
\,\bigg]
\,\bigg\}
\,.
\label{A2explicit}
\end{eqnarray}
There are no non-Abelian contributions to ${\cal{A}}_2$.\\[2mm]
{\it Step 2: Matching calculation.} --
The contributions to $C_1$ up to ${\cal{O}}(\alpha_s^2)$ are
determined by matching expression~(\ref{crosssectionexpansion})
directly to the
two-loop cross section calculated in full QCD in the (formal) limit
$\alpha_s\ll v\ll 1$ for stable quarks ($\Gamma_t=0$) keeping terms up
to $\alpha_s^2$ and NNLO in the expansion in $v$ and setting
$\mu_{\rm soft}=\mu_{\rm hard}$.\footnote{
For $\alpha_s\ll v\ll 1$, i.e.\ far away from the threshold regime, no
distinction between soft and hard scales is needed.
}
To obtain the contributions $\propto \alpha_s^2$ originating from the
potential $V_c$ in this limit we also
employ time-independent perturbation theory in analogy to
Eq.~(\ref{GreenfunctionBF}).  
The corresponding cross section in full QCD reads
($a_h\equiv\alpha_s(\mu_{\rm hard})$, $v\equiv (E/M_t)^{1/2}$)
\begin{eqnarray}
\lefteqn{
R_{\mbox{\tiny 2loop QCD}}^{\mbox{\tiny NNLO}} \, = \,
N_c\,Q_t^2\,\bigg\{\,\bigg[\,
\frac{3}{2}\,v-\frac{17}{16}\,v^3
\,\bigg] +
\frac{C_F\,a_h}{\pi}\,\bigg[\,
\frac{3\,\pi^2}{4}-6\,v+\frac{\pi^2}{2}\,v^2
\,\bigg]
}
\nonumber\\[2mm] & & 
+\, a_h^2\,\bigg[\,
\frac{C_F^2\,\pi^2}{8\, v} 
+ \frac{3}{2}C_F\,\bigg(\,
- 2 \,C_F 
+ C_A \,\Big(\,
    -\frac{11}{24}\ln\frac{4\,v^2\,M_t^2}{\mu_{\rm hard}^2}+\frac{31}{72}  
\,\Big)
+ T\,n_l \,\Big(\, 
    \frac{1}{6}\ln\frac{4\,v^2\,M_t^2}{\mu_{\rm hard}^2}-\frac{5}{18}  
\,\Big)
\,\bigg) 
\nonumber\\[2mm] & & \qquad
+ \bigg(\, 
\frac{49\,C_F^2\,\pi^2}{192}  
  + \frac{3}{2}\,\kappa 
  +  \frac{C_F}{\pi^2}\, \Big( \frac{11}{2}\, C_A - 2\, T\,n_l \Big)
        \, \ln\frac{M_t^2}{\mu_{\rm hard}^2} 
  - C_F\Big( C_F+\frac{3}{2}\, C_A \Big)\,\ln v
\,\bigg)\,v
\,\bigg]
\,\bigg\}
\,,
\label{Rtwoloop}
\end{eqnarray}
where 
\begin{eqnarray}
\kappa & = &
C_F^2\,\bigg[\, \frac{1}{\pi^2}\,\bigg(\,
\frac{39}{4}-\zeta_3 \,\bigg) +
\frac{4}{3} \ln 2 - \frac{35}{18}
\,\bigg] 
- C_A\,C_F\,\bigg[\,  \frac{1}{\pi^2} \,\bigg(
\frac{151}{36} + \frac{13}{2} \zeta_3 \,\bigg) +
\frac{8}{3} \ln 2 - \frac{179}{72} \,\bigg] 
\nonumber\\[2mm] & & +
C_F\,T\,\bigg[\,
\frac{4}{9}\,\bigg(\, \frac{11}{\pi^2} - 1\,\bigg)
\,\bigg] +
C_F\,T\,n_l\,\bigg[\, \frac{11}{9\,\pi^2} \,\bigg] 
\,.
\label{kappadef}
\end{eqnarray}
The Born and ${\cal{O}}(\alpha_s)$~\cite{Kallensabry1} contributions
are standard.
At order $\alpha_s^2$ the contributions in Eq.~(\ref{Rtwoloop})
proportional to $C_F^2$, $C_A C_F$, $C_F T n_l$ and $C_F T$ have been calculated
in~\cite{Hoang4}, \cite{Carnecki1}, \cite{Hoang5,Voloshin1} and
\cite{Hoang5,Karshenboim1}, respectively.
The result for $C_1$ reads 
\begin{equation}
C_1 \, = \, 
1 - 4\, C_F\frac{a_h}{\pi} + 
a_h^2\,
\bigg[\,
\kappa 
+ \frac{C_F}{\pi^2}\,\bigg(\,
\frac{11}{3}\,C_A - \frac{4}{3}\,T\,n_l
\,\bigg)\,\ln\frac{M_t^2}{\mu_{\rm hard}^2}
+ C_F\,\bigg(\,
\frac{1}{3}\,C_F + \frac{1}{2}\,C_A
\,\bigg)\,
\ln \frac{M_t^2}{\mu_{\rm fac}^2}
\,\bigg]
\,.
\label{C1explicit}
\end{equation}
The consistency of the ``direct matching'' procedure ensures that
$C_1$ does not contain any energy-dependent terms. We would like to point
out that the factorization scale dependence of $\mbox{Im} [{\cal{A}}_1]$
is cancelled by the factorization scale dependence in $C_1$ up to a
small term $\propto C_F\,\alpha_s \frac{\Gamma_t}{M_t}
\ln(\frac{M_t}{\mu_{\rm fac}})$ (see also~\cite{Hoang3}). This term
remains as a consequence
of our ignorance of a consistent treatment of the finite width effects
at the NNLO level. Due to the small size of this contribution, however,
the corresponding ambiguity can be ignored for the examination of the
relativistic NNLO corrections determined in this paper.
The LO cross section can be recovered from the NNLO one in
Eq.~(\ref{crosssectionexpansion}) by taking into account only the
dominant contributions in the third line of Eq.~(\ref{A1explicit}) and
setting $C_1=1$, $C_2=0$, whereas the NLO cross section can be
obtained by incorporating also the ${\cal{O}}(\alpha_s)$ corrections
to the Coulomb potential and to the constant $C_1$. 
\begin{figure}[t!] 
\begin{center}
\leavevmode
\epsfxsize=3.4cm
\epsffile[220 420 420 550]{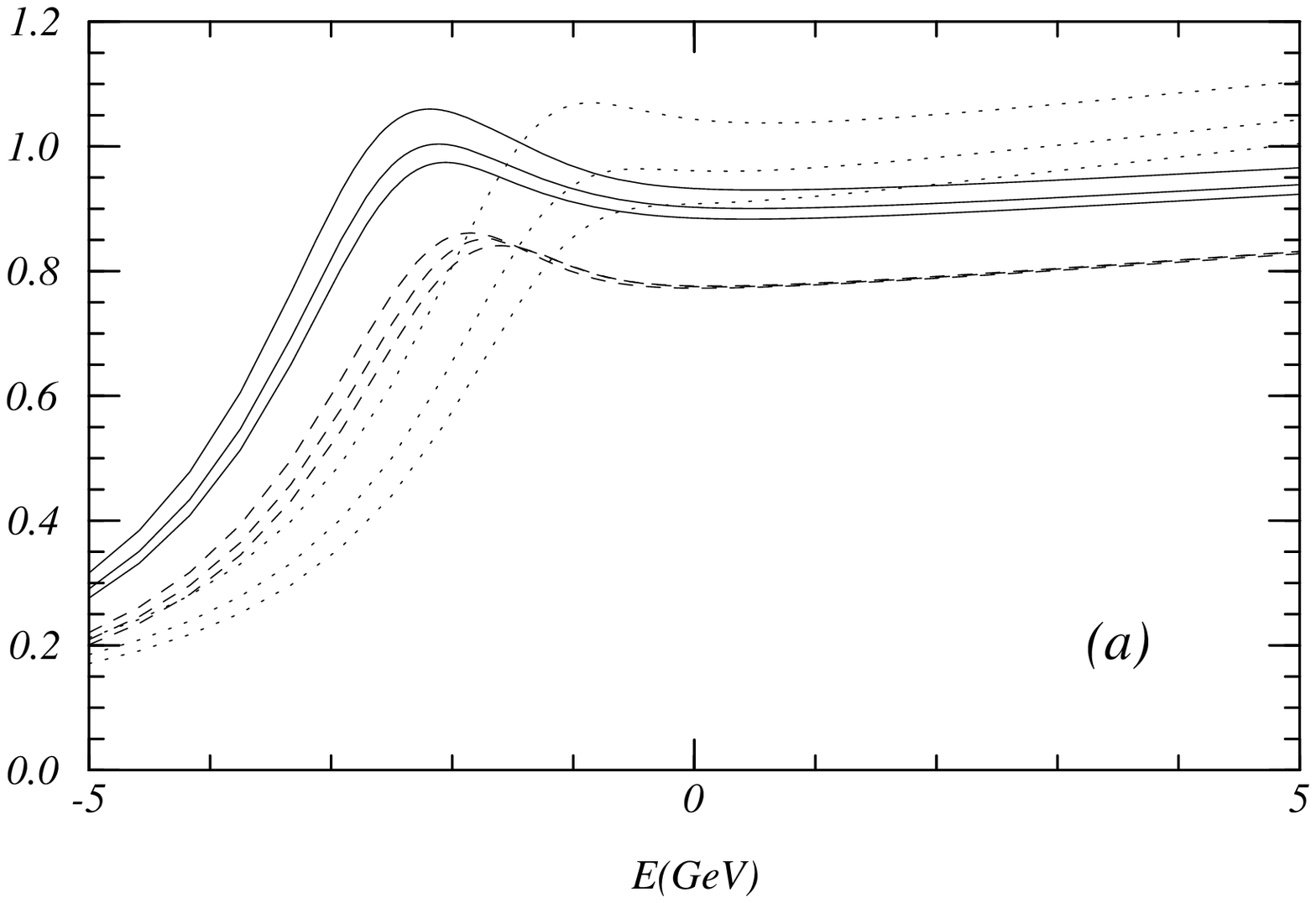}
\hspace{4.7cm}
\epsfxsize=3.4cm
\leavevmode
\epsffile[220 420 420 550]{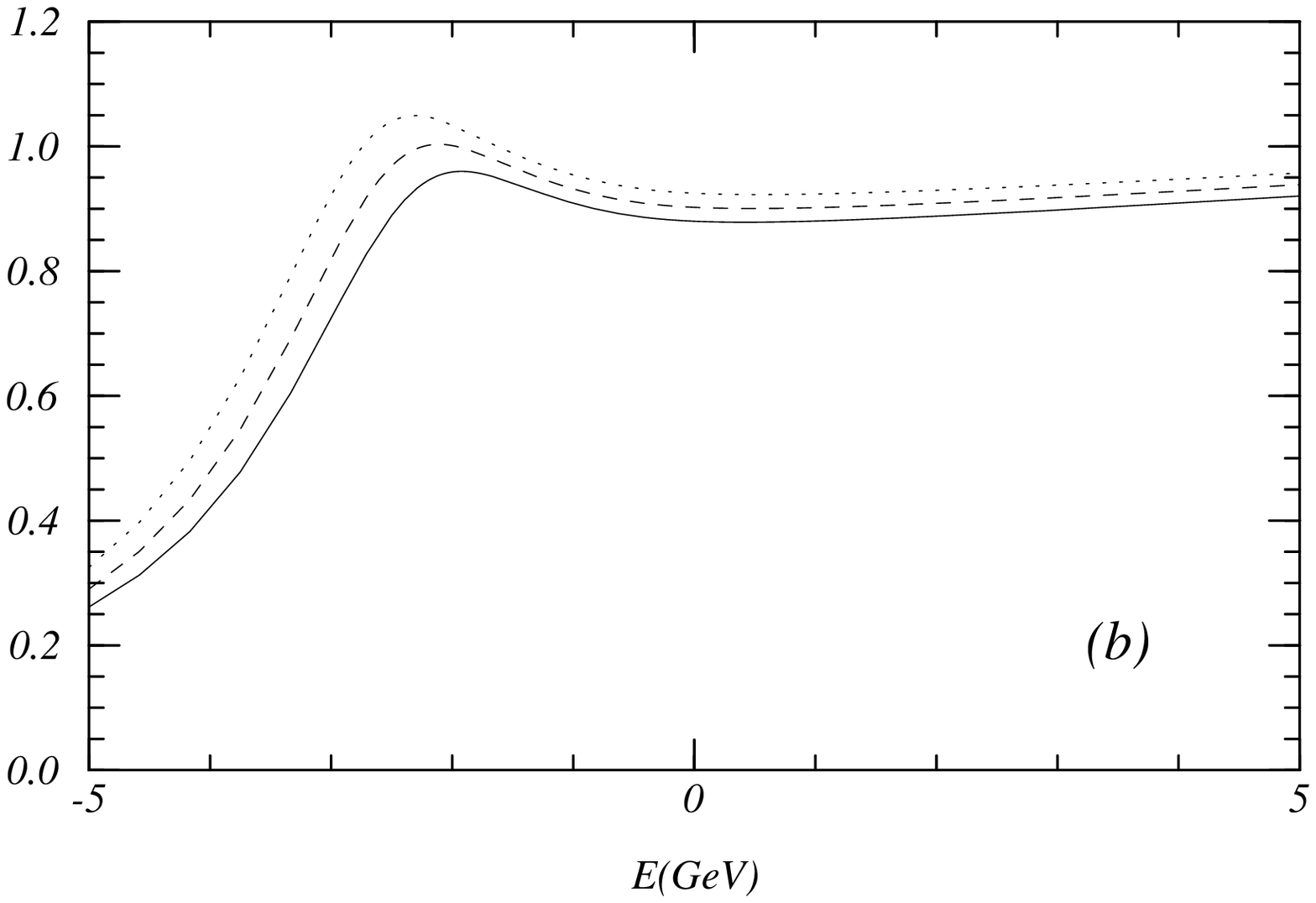}
%
%
\vskip  2.7cm
 \caption{\label{fig1} 
(a) The total normalized photon-mediated $t\bar t$ cross section at
LO (dotted lines), NLO (dashed lines) and NNLO (solid lines)
for the soft scales $\mu_{\rm soft}=50$ (upper lines), $75$
and $100$~GeV (lower lines). 
(b) The NNLO cross section for $\alpha_s(M_z)=0.115$ (solid line),
$0.118$ (dashed line) and $0.121$ (dotted line). More details and the
other parameters are given in the text.
}
 \end{center}
\end{figure}
In Fig.~\ref{fig1}(a) the LO (dotted lines), NLO (dashed lines) and NNLO
(solid lines) normalized cross sections are plotted versus $E$ in the range 
$-5$~GeV$\,< E < 5$~GeV for $M_t=175$~GeV, $\alpha_s(M_z)=0.118$ and
$\Gamma_t=1.43$~GeV. For the scales the choices 
$\mu_{\rm soft} = 50$ (upper lines), $75$ and $100$~GeV (lower lines) and
$\mu_{\rm hard} = \mu_{\rm fac} = M_t$ have been
made and two-loop running of the strong coupling has been used.
It is evident that the NNLO corrections are large. Compared to the NLO
cross section, the $1S$ peak is shifted towards smaller energies by
several hundred MeV and the large negative NLO corrections for
positive energies are compensated to some extent. Whereas the
location of the $1S$ peak is quite insensitive to changes in the soft
scale, the residual dependence of the normalization of the NNLO cross
section on the soft scale $\mu_{\rm soft}$ is not improved at all
compared to the NLO cross section. In fact, for energies above the
$1S$ peak it is worse for the NNLO cross section than for the NLO
one. The dependence of the NNLO cross
section  on the hard scale $\mu_{\rm hard}$ and the factorization
scale $\mu_{\rm fac}$ are, on the other hand, much smaller and,
therefore, not displayed here. The behavior of the NNLO corrections
clearly indicates that the convergence of the perturbative series for
the $t\bar t$ cross section is much worse than expected from the
general arguments given by Fadin and Khoze~\cite{Fadin1}.
For the normalization of the cross section we estimate, at least at
the present stage, a theoretical uncertainty at the level of five to ten
percent.
For comparison, in Fig.~\ref{fig1}(b)
the NNLO cross section is displayed
for $\alpha_s(M_z)=0.115$ (solid line), $0.118$ (dashed line) and
$0.121$ (dotted line) and $\mu_{\rm soft}=75$~GeV. The other
parameters are chosen as before. A more detailed examination of the
NNLO contributions will be carried out in a future publication.

This work is supported in part by the U.S.~Department of Energy under
contract No.~DOE~DE-FG03-90ER40546.

\sloppy
\raggedright
\def\app#1#2#3{{\it Act. Phys. Pol. }{\bf B #1} (#2) #3}
\def\apa#1#2#3{{\it Act. Phys. Austr.}{\bf #1} (#2) #3}
\def\lhc{Proc. LHC Workshop, CERN 90-10}
\def\npb#1#2#3{{\it Nucl. Phys. }{\bf B #1} (#2) #3}
\def\nP#1#2#3{{\it Nucl. Phys. }{\bf #1} (#2) #3}
\def\plb#1#2#3{{\it Phys. Lett. }{\bf B #1} (#2) #3}
\def\prd#1#2#3{{\it Phys. Rev. }{\bf D #1} (#2) #3}
\def\pra#1#2#3{{\it Phys. Rev. }{\bf A #1} (#2) #3}
\def\pR#1#2#3{{\it Phys. Rev. }{\bf #1} (#2) #3}
\def\prl#1#2#3{{\it Phys. Rev. Lett. }{\bf #1} (#2) #3}
\def\prc#1#2#3{{\it Phys. Reports }{\bf #1} (#2) #3}
\def\cpc#1#2#3{{\it Comp. Phys. Commun. }{\bf #1} (#2) #3}
\def\nim#1#2#3{{\it Nucl. Inst. Meth. }{\bf #1} (#2) #3}
\def\pr#1#2#3{{\it Phys. Reports }{\bf #1} (#2) #3}
\def\sovnp#1#2#3{{\it Sov. J. Nucl. Phys. }{\bf #1} (#2) #3}
\def\sovpJ#1#2#3{{\it Sov. Phys. LETP Lett. }{\bf #1} (#2) #3}
\def\jl#1#2#3{{\it JETP Lett. }{\bf #1} (#2) #3}
\def\jet#1#2#3{{\it JETP Lett. }{\bf #1} (#2) #3}
\def\zpc#1#2#3{{\it Z. Phys. }{\bf C #1} (#2) #3}
\def\ptp#1#2#3{{\it Prog.~Theor.~Phys.~}{\bf #1} (#2) #3}
\def\nca#1#2#3{{\it Nuovo~Cim.~}{\bf #1A} (#2) #3}
\def\ap#1#2#3{{\it Ann. Phys. }{\bf #1} (#2) #3}
\def\hpa#1#2#3{{\it Helv. Phys. Acta }{\bf #1} (#2) #3}
\def\ijmpA#1#2#3{{\it Int. J. Mod. Phys. }{\bf A #1} (#2) #3}
\def\ZETF#1#2#3{{\it Pis'ma Zh. Eksp. Teor. Fiz. }{\bf #1} (#2) #3}
\def\jmp#1#2#3{{\it J. Math. Phys. }{\bf #1} (#2) #3}
\def\yf#1#2#3{{\it Yad. Fiz. }{\bf #1} (#2) #3}

\end{document}